\documentstyle[12pt,epsf]{article}
\input epsf.sty
\epsfverbosetrue

\begin{document}
\title{The Geometry of Stochastic Reduction of an Entangled System} 
\author{Ari Belenkiy$^a$, Steve Shnider$^a$, Lawrence Horwitz$^{b,c}$ }

\date{{\small
a) Department of Mathematics, Bar-Ilan University, Ramat Gan, Israel;
 b) School of Physics, Tel Aviv University, Ramat Aviv, Israel;
 c) Department of Physics, College of Judea and Samaria, Ariel, Israel}}

\maketitle


\par
 

\abstract{{\small\noindent  
We show that the method of stochastic reduction of linear 
superpositions can be applied to the process of disentanglement for the 
spin-$0$ state of two spin-$\frac12$ particles. We describe the geometry of 
this process in the framework of the complex projective space.}}

\section{Introduction}

A pure quantum state of a system is  a vector in a Hilbert space, which may be 
represented as a linear combination of a basis of eigenstates of an observable (self-adjoint operator) or of several commuting observables. 
Let us suppose that the eigenvalues corresponding to the eigenstates 
of the Hamiltonian operator of a system are the physical quantities 
measured in an experiment. If the action of the 
experiment is modelled by a dynamical interaction induced by a term in the 
Hamiltonian of the system, and its effect is computed by means
of the standard evolution according to the Schr\"odinger equation, the final 
state would retain the structure of the original linear superposition. One
observes, however, that the experiment provides a final state that is one 
of the basis eigenstates and the superposition has  been destroyed. 
The resulting
process is called reduction or collapse of the wave function. The history of 
attempts to find a systematic framework for the description of this process 
goes back very far in the development of quantum theory (e.g., the problem of 
Schr\"odinger's cat \cite{sch}). In recent years significant progress has been 
made. Rather than invoking some random interaction with the environment and 
attributing the observed decoherence, i.e. collapsing of a linear superposition, 
to the onset of some uncontrollable phase relation, more rigorous methods have
 been developed, which add to the Schr\"odinger equation stochastic terms 
corresponding to Brownian fluctuations of the wave function. Since a pure
quantum state of a system corresponds to an equivalence class of vectors 
modulo scaling by a non-zero complex number, corresponding to the norm and an 
overall phase factor \cite{wig, mac}, it is natural to develop models for 
collapse in the setting of a projective space \cite{k, bh1}. 
Associated to an $N$-dimensional complex Hilbert space, we have the projective
space ${\bf CP}^{n-1}$ equipped with the canonical Fubini-Study metric. 

In this paper, we shall apply some of these methods of state reduction to the
phenomena considered in the famous paper by Einstein-Podolsky-Rosen \cite{epr}
explored experimentally by Aspect \cite{asp}, and analyzed by Bell for its 
profound implications in quantum theory \cite{be1, be2}. The system to be 
studied consists
of a two particle quantum state, where each particle has spin $\frac12$. 
The two body state of total spin zero has the special property known as 
``entangled" for which  a determination of the state of one particle implies 
with certainty the state of  the second. The problems recognized by  EPR and 
studied extensively by Bell arise when the two entangled particles are very 
far apart. 
  
The states of the two particle system which we shall consider are the  
equivalence classes of vectors in the tensor product of two spin $\frac12$ 
representation spaces ${\cal H}\otimes {\cal H}$, where ${\cal H}$ corresponds
 to the states of one of the constituents. We shall describe the experimental detection of the entangled states in terms of mathematical models recently developed for describing the reduction, or collapse, of the wave function. One begins with an entangled state, corresponding to the $1$-dimensional spin $0$ representation with basis vector the linear superposition:
\begin{equation}
\label{1}
|s=0\rangle:=\frac1{\sqrt 2}(|\uparrow \rangle_1\otimes|\downarrow \rangle_2
-|\downarrow \rangle_1\otimes |\uparrow\rangle_2.
\end{equation}
Here $1,2$ refer to the two spin $\frac12$ representations, each one with a 
basis $\{|\uparrow\rangle$, $|\downarrow\rangle\}$, corresponding to spin up 
and spin down, resp., relative to an arbitrary but fixed  direction.  
The full tensor product representation is a sum of 
this spin $0$ representation and a complementary spin $1$ representation.

The first stage of reduction, using  the  stochastic evolution model 
developed by 
Diosi, Ghirardi, Pearle, Rimini, Brody and Hughston \cite{bh1, di, hu, gpr}, 
and references therein, gives rise to a density matrix, a linear combination 
of projections on disentangled states with Born probability coefficients. 
The second stage of reduction is the detection of the configuration of 
disentangled states, which we will not discuss in detail here.  
Assume that one initially has an entangled spin $0$ state of a two particle 
system and then by some physical process the two particles become separated 
and far apart. 
Measurement of the first particle in the spin down state then implies with 
certainty that the second particle is in the spin up state, measured in the 
same direction. For the spin $0$ state this direction is arbitrary.
The question is often raised as to how the state of the second particle can 
respond to the arbitrary choice of direction in the measurement of the first. 
This question is dealt with here  by the addition of an additional term to the
 Hamiltonian, which we attribute to the presence of the measurement apparatus.
 On this basis, we shall attempt here to give a mathematical description of  
the process underlying such a measurement.

The state $|s=0\rangle$ is represented  in equation (\ref{1}) as a linear 
superposition. As noted above, recently developed methods for describing state
 reduction can account for a reduction of this superposition to one or the 
other of the product states occurring on the right hand side of eq. 
(\ref{1}) in a simple way if these states are eigenstates of the self-adjoint 
infinitesimal generator (Hamiltonian) of the evolution.

Suppose, for example, that the Hamiltonian has the form,
\begin{equation}\label{2}
H=H_0+H_1
\end{equation}
where $H_0$ contains the spin-independent kinetic energy of the two particles,
\begin{equation}\label{3}
H_0= p_1^2/2m_1 + p_2^2/2m_2,
\end{equation}
describing the free motion, but $H_1$ has the special form
\begin{eqnarray}\label{4}
H_1&=&\sum \lambda_{i,j} P_{i,j}\\\
&=&\sum \lambda_{i,j}(|v_i\rangle_1\otimes |v_j\rangle_2)\otimes
(_1\langle v_i|\otimes _2\langle v_j|),\nonumber
\end{eqnarray}
where the sum is over $i,j=1,2$ and $v_1=\uparrow, v_2=\downarrow$. 
We show in the next section that, applying  the method of adding  
a Brownian term to the Schr\"odinger equation, \cite{hu, gpr, di}, 
causes the system to evolve  into one or the other of
the eigenstates $|v_i\rangle_1\otimes |v_j\rangle_2$ with the correct
Born {\it a priori} probabilities \cite{ah, gpr}. In the case of
an initial state of the form (\ref{1}), the resulting asymptotic state is
either $|\uparrow\rangle_1\otimes |\downarrow\rangle_2)$ or 
$|\downarrow\rangle_1\otimes |\uparrow\rangle_2)$,
each with probablity $\frac12$. Such a configuration is called a mixed state.

We should remark that if the two particles correspond to identical fermions, 
then indices $1,2$ are basically indistinguishable and the two states 
$|\uparrow\rangle_1\otimes |\downarrow\rangle_2)$ and 

\noindent
$|\downarrow\rangle_1\otimes |\uparrow\rangle_2)$
should appear with equal weights. However, since the particles are located
far apart when the measurement takes place, there is no overlap of the one
particle wave functions, and the Fermi antisymmetry is not required.
Thus the presence of two widely separated detectors can split the degeneracy 
into distinct states, which can, in fact, imply that $\lambda_{1,2}\neq \lambda_{2,1}$.

The second stage of reduction, as pointed out above,  corresponds to the 
destruction of the two body state by one-particle filters. 
The state actually measured is
a ``separated system" of two particles. We assume that the two filters,
which we denote $M_u$ and $M_d$  have the property that if the state has 
the form $|\uparrow\rangle_1\otimes |\downarrow\rangle_2$, then $M_u$ applied 
to particle $1$ and $M_d$ applied to particle $2$ succeed with certainty. 
We shall not discuss the extensive literature dealing with the problem of 
representing separated systems \cite{ae,pi}. We take as our primary task 
the description of the first stage of this reduction process.

In the application of the technique of state reduction, it is usually 
assumed that the evolution  is governed by the physical
nature of the system before the measurement process. However, in an undisturbed
quantum system the linear supposition of states evolves according to a one
parameter group of unitary operators which preserves the superposition and for
which there is no collapse. One may understand the Brownian fluctuations 
leading
to collapse as induced by the presence of measurement apparatus. In the same
way, the component $H_1$ of the Hamiltonian may be thought of as induced by
the measurement apparatus, which, in our formulation of the problem, 
disentangles the states, even to the extent of defining the orientations for 
the states $|\uparrow\rangle, |\downarrow\rangle$. 

In terms of the projective geometry the disentangled states lie in a quadric
which is naturally defined by the identification of the underlying 
$4$-dimensional complex vector space as the tensor product of two 
$2$-dimensional complex vector spaces. The entangled state $|s=0\rangle$ lies 
outside this quadric, and the stochastic evolution of the system moves the 
point $|s=0\rangle$ into the quadric in the first stage of reduction.

In the next section we review the geometric approach to quantum mechanics
in terms of projective space and describe the geometry of entanglement. In 
the following sections we show how the introduction of the modified 
Hamiltonian in Hughston's model for stochastic evolution gives a theoretical 
framework for describing Aspect's experiments. 

\section{Geometric quantum mechanics}
We begin with a quick review of the geometric framework for quantum mechanics
in terms of Hamiltonian symplectic dynamics on the quantum mechanical state 
space introduced by Kibble [1] and developed further by Brody, Hughston and 
others. For simplicity we will assume that for each time $t$, the wave 
function $\psi(x,t)$ belongs to a fixed  finite dimensional complex Hilbert 
space and is represented as a linear superposition of a finite basis of states
 $\psi_j$
$$ \psi = z^1 \psi_1 + z^2 \psi_2 +...+ z^n \psi_n. $$
The normalization condition demands that 
$$ |z^1|^2 + |z^2|^2 +...+ |z^n|^2 = 1,$$
and since wave functions related by a phase factor $e^{i\alpha}$ represent the
 same physical state, the time evolution of the system is actually taking place in complex projective $n-1$-space
$$S^{2n-1}/ S^1\equiv {\bf CP}^{n-1}.$$
\par\bigskip
The space ${\bf CP}^{n-1}$ is the set of  equivalence classes of complex  $n$-tuples
modulo multiplication by a non-zero complex number. An equivalence class
is represented by $(z^1: ...\ z^j \ ... :z^n),$ and the $z^i$ are called 
the {\it homogeneous} coordinates of that point. 
The eigenstate $\psi_j$ corresponds to the point 
$z^j= (0: ... \ 1_j ... \ :0)$.    
\par\bigskip
The time evolution of the quantum state is given by the Schr\"odinger equation on ${\bf C}^n$:
$$i d z^j/dt = H_{k,j} z^k,$$
with $H_{k,j} = (\psi_k, \bf H \psi_j)$.
In a coordinate patch of ${\bf CP}^{n-1}$, for example, $z^n\neq 0$,
 with coordinates $\{x^a| a=1,...,2(n-1)\}$,
$ x^a + i x^{a+n}:=z^a/z^n$ the Schr\"odinger equation 
can be expressed in Hamiltonian form
\begin{equation}
\hbar dx^a/dt = 2 \Omega^{ab} \nabla_b H(x) ,
\label{Schrod}
\end{equation}
where $\nabla$ is a covariant derivative on $CP^{n-1}$ with a connection form 
associated with Fubini-Studi metric, $\Omega^{ab}$ is the
symplectic structure  and the real-valued function 
(observable) $H(x)$, is defined by
\begin{equation} 
 H(x) = 
{\Sigma H_{j,k} z^j \ \bar z^k \over {\Sigma |z^j|^2}}.
\label{perfectfunction}
\end{equation}
If the operator {\bf H} is diagonal in the representation provided by 
$\{\psi_j\}$, e.g., with eigenvalues $\lambda_j$, $H(x)$ takes the form  
$$ H(x) = {\Sigma_j \lambda_j |z^j|^2 \over \Sigma |z^j|^2}$$
which is a function with critical points at $z^j=(0: \ 1_j: \ 0)$. 
\par\bigskip
The projective space geometry naturally lends itself to the computation of
transition probabilities. The transition probability 
from state $X$ to state $Y$ is given by
\begin{equation}
Prob(X,Y) = {\langle X|Y\rangle \langle Y|X\rangle \over \langle X|X\rangle 
\langle Y|Y\rangle},
\label{scalar}
\end{equation}
which has a simple relation to  the geodesic distance  with respect to the 
Fubini-Study 
metric between  $X$ and $Y$ considered as points in ${\bf CP}^{n-1}$.
 Calling this distance $\theta$, we have, \cite{hu},
$$cos^2({\theta\over 2})={\langle X|Y\rangle \langle Y|X\rangle \over  \langle X|X\rangle \langle Y|Y\rangle}.$$
This, in particular, means that two {\it conjugate} or {\it orthogonal} 
points have geodesic distance $\pi$ between them.

The state space for a pair of spin-${1 \over 2}$ particles is the projective
space of ${\bf C}^2\otimes {\bf C}^2$, which we indentify with ${\bf CP}^3$. 
We represent the basis of ${\bf C}^2$ as $\uparrow, \downarrow$, and the
basis ${\bf C}^2\otimes {\bf C}^2$ as $ \uparrow\otimes  \downarrow, 
\uparrow\otimes \uparrow, \downarrow\otimes  \downarrow,  \downarrow\otimes \uparrow.$ 
Let $(x:y:z:w)$ be the homogeneous coordinates  corresponding to this basis.  
\par\bigskip
The {\it singlet} state (total spin-0 case)is represented in homogeneous coordinates as  
$$ P_0 = (1:0:0:-1)$$
(it is also represented by the line in $C^4$ with $ x = -w, y = z = 0$).
\par\bigskip
The {\it triplet} representation is the orthogonal hyperplane $L$,
whose equation in   homogeneous coordinates is
$$ L = \{x-w=0 \}\quad\mbox{\rm or, in parametric form}\quad L = \{(x:y:z:x)\}.$$
Let us describe the space of possible representations of the eigenstates  of 
the {\it spin-z} operator. The directions of the $z$-axes of
a system of two particles are parametrized by $CP^1 \times CP^1$.
\par\bigskip 
The manifold of such states is imbedded in our $CP^3$ as the decomposable $2$-tensors,
$(a\uparrow +b\downarrow)\otimes (c\uparrow+ d\downarrow)$:
which gives the {\it Veronese} embedding 
$$((a:b),(c:d))\mapsto (ad: ac: bd: bc)$$ 
 of $CP^1 \times CP^1$ onto the quadric represented by the equation
\begin{equation}
\label{quadric}
Q = \{xw=yz\}.
\end{equation}
The quadric $Q$ intersects the plane $L$ in a conic :
\begin{equation}
\label{conic}
C = \{x^2=yz,x=w\}.
\end{equation}
The point
$$  P_{\uparrow \uparrow} = (0:1:0:0)$$  on the conic corresponds to the 
initial spin axis. 
The point $$P_{\downarrow \downarrow} = (0:0:1:0)$$ is the unique point in 
the conic (\ref{conic}) which is conjugate (orthogonal) to
$P_{\uparrow \uparrow} = (0:1:0:0)$ relative to the standard Hermitian inner 
product.
So far we constructed only two eigenstates of the {\it spin-z} operator.
The third triplet state $P_1$ of the 
spin operator  lies at the intersection of the tangents to the conic at 
$P_{\uparrow \uparrow}$ and $P_{\downarrow \downarrow}$, see \cite{hu},
and is given by the equations $y=z=0, x=w$: 
$$ P_1 = (1:0:0:1) .$$ 

A basis for the $0$ eigenstates of the {\it spin-z} operator 
in the full four dimensional representation is given by the intersection of the line
$$\overline{P_0 P_1} = (\mu+\nu: 0: 0 :\mu - \nu)$$ with the  
the quadric $\{xw = yz\}$ in the  two distinct points with $\mu = \pm \nu$:
$$  P_{\uparrow \downarrow} = (0:0:0:1) $$ 
and 
$$  P_{\downarrow \uparrow} = (1:0:0:0) .$$ 
In this framework, we have constructed the geometry of four spin states 
spanning ${\bf C}^2\otimes {\bf C}^2$. Moreover, we have explained that 
disentangled states form a quadric in the associated projective space, and 
that the spin $0$ entangled state, lying outside this quadric is a 
distinguished point.  

\section{Collapse of the entangled state}

We now describe the mechanism by which an initial entangled state, 
corresponding to this distinguished point, can evolve into a disentangled 
state in the quadric. To see how this occurs, we review briefly the mechanism 
of wave function collapse induced by stochastic fluctuations of the 
Schr\"{o}dinger evolution. We follow closely the method of Hughston 
\cite{hu} (see also \cite{abbh,ah}).

\par\bigskip
In the stochastic reduction model of Hughston the system is governed 
by the following  stochastic differential equation:
\begin{equation}
\label{twiddle}
 d x^a = (2\Omega^{a,b} \nabla_b H -  {1 \over 4} \sigma^2 \nabla^a V) dt
+ \sigma \nabla^a H d W_t
\end{equation}
where $$ V(x) =  \nabla_a H(x) \nabla^a H(x)$$ 
is a so-called {\it quantum uncertainty}.(Where it is not mentioned 
explicitly, the indexes are lifted by the metric.) 

>From It\^ {o} theory it immediately follows that above process has two 
basic properties:
\par\bigskip
1) Conservation of Energy
$$ H(x_t) = H(x_0) + \sigma \int_0^t V(t) dW_t$$

2) Stochastic reduction
\begin{equation}\label{star}
 dV = - \sigma^2 V(x_t)^2 dt + 
\sigma \nabla_x V(x_t) \nabla^x \beta(x_t) dW_t
\end{equation}
where  $$\beta(x) = \nabla_a H(x) \nabla^a V(x)$$ is the ``third'' moment.

It follows from (\ref{star}) that the expectation $E[V]$ of the stochastic process
obeys the relation \cite{hu, ah}
$$ E[V_t]= E[V_0] - \sigma^2 \int ^t_0 ds E[V_s^2],$$
and since
$0\leq E[(V_s-E[V_s])^2]=E[V_s^2]-(E[V_s])^2,$
$$E[V_t]\leq E[V_0] - \sigma^2 \int ^t_0 ds E[V_s]^2.$$
Since $V_s$ is positive, this implies that $E[V_\infty]=0$,
and (up to measure $0$ fluctuations) $V_t\rightarrow 0$ as $t$ tends to $\infty$.
Since 
$$ V=\langle \psi, (H -\langle H\rangle)^2 \psi)/||\psi||^2,$$
where $\langle H\rangle=E[H]=\langle \psi,H\psi\rangle/||\psi||^2$,
$$V=0 \quad\mbox{\rm implies}\quad ||(H-\langle H \rangle)\psi||=0,$$
 and 
$H\psi=\langle H\rangle \psi$, so $\psi$ is an eigenvector of the 
Hamiltonian. 

Note that the system we have described  brings the system to one or another 
of the eigenstates of the Hamiltonian $H$, with the Born probability given by
the initial state, \cite{ah, gpr}. Therefore, the final configuration 
corresponds to a mixed state, with each component an eigenstate of $H$. 

We now apply the mechanism to what we have called the first stage of the
Aspect type experiment \cite{asp}, the evolution from an entangled state to 
a disentangled state. 

Let us suppose that the system of two spin $\frac12$ particles is initially 
in  the entangled spin $0$ state, and the two particles move away from
each other, according to the motion  generated by (\ref{3}). As the particles 
approach some neighborhood of the detector, the Schr\"odinger evolution, 
$$i {{\partial \psi}\over{\partial t}}= H_0\psi$$
is modified by the Brownian fluctuations appearing in (\ref{twiddle}), 
presumably induced by the detectors and their interactions, 
for example, with a set of quantum fields. 
We suppose, as well, that the filters of the apparatus induce  a self-adjoint
perturbation $H_1$ of the Hamiltonian itself, so that the system evolves, as 
in (\ref{2}), according to a perturbed Hamiltonian $H=H_0+H_1$ in addition 
to the effect of the Brownian fluctuations. In order that the quantum state 
converge by stochastic reduction to one of the disentangled states, 
$ \uparrow\otimes  \downarrow, \uparrow\otimes \uparrow
\downarrow\otimes  \downarrow,  \downarrow\otimes \uparrow$, we suppose the perturbation to
be of the form (\ref{4}). The component  $H_0$ of the Hamiltonian induces an 
irremovable dispersion, but the residual dispersion can be as small as we wish.

As we have pointed out, for identical particles there may be a degeneracy  between the
states $\uparrow\otimes \downarrow$ and $\downarrow\otimes \uparrow$; since the
filters in the experiment are arranged in one or the other of these configurations, one
expects this degeneracy to be broken. Since one particle has moved in one direction
and the second in another (with eventually no overlap of the wave functions), the particles
then become effectively distinguishable and the induced Hamiltonian is not required to be
degenerate. Therefore the  final state may become disentangled, as we noted in the Introduction. 

The evolution (\ref{twiddle}) corresponds to the motion of  the point in ${\bf CP}^3$,
going from a singlet state to a limit point in the quadric, for example $\uparrow\otimes
\downarrow$ occurring with the corresponding Born probability.  As we have pointed out
the second stage of the detection, due to the direct action of the detector, must destroy
the two body state and create a state of a so-called ``separated system" in which one particle is seen with spin up and the other with spin
down in two separate (although essentially simultaneous) experiments.  The mathematical
framework for such separated systems is not completely clear \cite{ae, pi}.
As Aerts \cite{ae} has shown, the set of propositions of such a system is the direct sum of two lattices and does not correspond to a lattice of subspaces of a Hilbert space.
We assume however that the outcome of two measurements  corresponds to the configuration
of the two body state just before the measurement, an assumption generally made in applications of the quantum theory. 

Furthermore, we may ask about a situation in which the two filters are not oriented in
opposite directions, but at an angle to each other. In this case the  bases of the two
spin $\frac12$ representation would not correspond. One basis would be  $\uparrow, \downarrow$ and the other would be $\nwarrow, \searrow$ where 
$$\nwarrow=\cos(\theta/2)\uparrow +\sin(\theta/2) \downarrow,\quad
\searrow=-\sin(\theta/2)\uparrow +\cos(\theta/2)\downarrow $$
The computation of the Born probability from a singlet state to a 
final state determined by the filters, say of the form $\nwarrow\otimes \downarrow$ would
be $\cos^2(\theta/2)$ ( eq. (\ref{scalar})) in agreement with experiment. 
In this way, arrangements of the filters can effect perturbations of the 
Hamiltonian that can cause the system to evolve
to the appropriate point of the quadric of disentangled states.

\section{Some concluding remarks}
We have discussed a mechanism based on stochastic reduction, corresponding to a
particular class of irreversible processes, which models the evolution of an 
entangled two-body system to a disentangled state.  As an extension of this
idea, one may consider a problem with a natural degeneracy of some initial 
state for which the the presence of effective detectors of some type induces a
 perturbation in which stochastic reduction takes place, as in the asymptotic 
cluster decomposition of products of quantum fields reducing a $N$-body system
 to $M$ $k$-body systems or the formation of local correlations in $N$-body 
systems such as liquids, or spontaneous symmetry breaking. 
In all these cases, due to the existence of continuous spectra, there will be 
some residual dispersion in the final state, although possibly very small. We 
are currently studying possible applications of the methods discussed here
to such configurations.


\begin{thebibliography}{ABCD}

\bibitem{abbh} S. L. Adler, D. C. Brody,T. A. Brun, L. P. Hughston,{\em J. Phys. A}
{\bf 34} (2001) 8795.

\bibitem{ah} S. L. Adler, L. P. Horwitz, {\em J. Math. Phys.} {\bf 41} (2000) 
 2485, errata {\bf 42},(2001) 976.


\bibitem{asp} A. Aspect, Proposed experiment to test the nonseparability of quantum mechanics, {\em Phys. Rev. D}{\bf 14} (1976) 1944-51, see Wheeler, Zurek, loc. cit. pp. 435-442.

\bibitem{ae} D. Aerts, {\em Int. Jour. of Theor. Phys.} {\bf 38} (1999) 289.

\bibitem{be1} J. Bell, On the problem of hidden variables in quantum mechanics, 
{\em Rev. Mod. Phys.}{\bf 38} (1966) 447-452.

\bibitem{be2} ---, {\it Speakable and Unspeakable in Quantum Mechanics} Cambridge University Press 1987.

\bibitem{bo}  D. Bohm  {\it Quantum Theory}. Prentice Hall 1951.

\bibitem{bh1}D. C. Brody and L. P.  Hughston, Geometric quantum mechanics. 
{\em J. Geom. Phys.} {\bf 38} (2001), 19-53.

\bibitem{di} Diosi, {\em J. Phys. A} {\bf 21} (1988) p. 2885, {\em Phys. Lett.} {\bf 129A}
(1988) p. 419, {\em Phys. Lett.} {\bf 132A} (1988) p. 233.

\bibitem{dnb}Dubrovin B., Novikov S. and Fomenko A. {\it Modern Geometry}. 3 vols.
NY. Springer 1984.
 
\bibitem{epr} A. Einstein, B. Podolsky, N. Rosen, Can a quantum mechanical 
description of physical reality be considered complete, 
{\em Phys. Rev. } {\bf 47} (1935) 777-780,
see Wheeler, Zurek, loc. cit., 138-151.

\bibitem{gib} G.W. Gibbons, Typical states and density matrices. 
{\em J. Geom. Phys.} {\bf 8} (1992), 147-162.

\bibitem{gpr}G. C. Ghirardi, P. Pearl, A. Rimimi, {\em Phys. Rev. A} {\bf 42} (1990) 78.

\bibitem{hu}L. P. Hughston, Geometry of stochastic state vector reduction.
{\em  Proc. Royal Soc. Lond A} {\bf 452} (1996), 953-79.

\bibitem{k}T.W.B. Kibble, Geometrization of quantum mechanics.
{\em Commun. Math. Phys.} {\bf 65} (1979), 189-201.

\bibitem{mac} G. Mackey, Mathematical Foundations of Quantum Mechanics, 
Benjamin-Cummings Advanced Book Program, Reading, Mass., 1963.

\bibitem{pi} C. Piron, M\'ecanique Quantique, Bases and Applications,  Presses
Polytechniques et Univesitaires, Lausanne, (1990).

\bibitem{sch}E. Schr\"odinger,  The present situation in quantum mechanics, 
{\em Proc. Am. Phil. Soc.} {\bf 124} (1980) 323-38, translation of the original 
articles in {\em Naturwissenschaften} {\bf 23} (1935) 807-812, 823-828, 844-849,
see Wheeler, Zurek, loc. cit., pp. 152-168.

\bibitem{wz} Wheeler and W. H. Zurek, eds. {\em Quantum Theory and Measurement} Princeton U. Press (1983)

\bibitem{wig}E. Wigner, Interpretation of quantum mechanics, p. 260-315 in Wheeler, 
Zurek, loc. cit.
\end{thebibliography}
\end{document}